# Photonic Crystal Microring Resonators on a Hybrid Silicon Nitride-on-Lithium Niobate Platform

Zhongdi Peng, Rakesh Krishna, Xi Wu, Amir H. Hosseinnia, Tianren Fan, and Ali Adibi[1*]

[1]School of Electrical and Computer Engineering, Georgia Institute of Technology, Atlanta, Georgia 30318, USA
ali.adibi@ece.gatech.edu

**Photonic-crystal resonators (PhCRs) have been widely used in nonlinear integrated photonics for frequency engineering applications. A microwave-assisted frequency converter based on PhCRs highlights its precise control of frequency (enabled by creation of a pair of supermodes by a corrugated PhCR) and bidirectional frequency conversion. In this paper, we demonstrate a high-quality PhCR on a hybrid silicon nitride-on-lithium niobate-on-insulator (SiN-on-LNOI) platform for the first time for voltage-driven flexible frequency conversion using the electro-optic effect (0.85 pm/V). The fabricated PhCR has a large supermode splitting bandwidth = 14.6 GHz and an intrinsic quality factor (Q) = 1.47×10$^5$. Using different periodic corrugation amplitudes in the fabricated PhCRs enables the precise control of mode splitting with a ratio of 93.5 MHz/nm between the mode splitting bandwidth and the corrugation amplitude.**

Optical frequency conversion is drastically needed in microwave photonics [1,2], optical networking [3,4], and quantum computing [5,6]. Due to the large mode confinement and strong light-matter interaction, integrated photonic platforms are widely used for frequency conversion by leveraging the acousto-optic effect [7,8], spectral shearing [9,10], adiabatic tuning [11,12], and the electro-optic (EO) effect [13,14,15]. Acousto-optic modulators enable frequency shifts up to a few GHz [7], but the high conversion efficiency has only been posted in bulk materials. Spectral shearing using opto-mechanical devices shows large frequency shifts [10] but requires synchronization of the optical pulse and the RF signal. The adiabatic tuning of the optical cavity resonance drives the resonant frequency based on phase control with complex programmed periodic series and large power consumption [11,12]. The frequency conversion based on EO phase modulation typically generates additional sidebands with intrinsically low conversion efficiencies. Despite recent progress, the measured conversion efficiency is still limited in the reported single-sideband EO-modulated devices [13,14]. A frequency converter with bi-directional conversion and near-unity conversion efficiency in a coupled-resonator optical waveguide (CROW) device has been reported based on EO modulation [16]. Frequency conversion is facilitated by the creation of two supermodes through coupling two resonators with the same mode structure. However, the reported devices consume extra power for mode-matching between the coupled resonators to create symmetric split modes for maximum conversion efficiency. Moreover, the coupling gap between the two microresonators needs to be carefully designed for an appropriate coupling strength to precisely control the splitting bandwidth between the two supermodes to increase the flexibility in achieving a desired frequency conversion.

Here we introduce a new structure for controlled mode-splitting in a photonic-crystal resonator (PhCR). In conventional microring resonators, clockwise (CW)- and counterclockwise (CCW) modes are frequency degenerate. By adding periodic corrugation to the periphery of the microring, we form a PhCR with the desired level of coupling between CW and CCW modes. This lifts the frequency degeneracy to create a pair of supermodes with different frequencies and azimuthal mode distributions (Fig. 1(a)). The mode splitting bandwidth $\beta_m$ varies linearly with the geometric corrugation amplitude $A_n$ in a PhCR with azimuthal mode number $m$ and a modulation of radius of $r = r_0 + A_n \cos(2n\phi)$, where $r_0$ is the original radius and n is the number of corrugation periods [17,18]:

$$\beta_m = k\omega_m A_n, \quad (1) \quad \text{with}$$

$$k = \frac{\int dS \left[ (\epsilon_{SiN}-1)\epsilon_0 |E_{m,r}(\phi)|^2 + \left(1 - \frac{1}{\epsilon_{SiN}}\right)\frac{1}{\epsilon_0}|D_{m,z}(\phi)|^2 \right] \cos^2(m\phi)\cos(n\phi)}{\int dV \epsilon (|E_{m,r}(\phi)|^2 + |E_{m,z}(\phi)|^2)}. \quad (2)$$

Here, $E_{m,r}$ and $D_{m,z}$ are the electric field and electric displacement field parallel to and perpendicular to the corrugation boundary $dS$, respectively. $\phi$ is the angle along the periphery of the PhCR, $\omega_m$ is the unperturbed resonant frequency of the mode with azimuthal mode number $m$, $\epsilon_0$ is the vacuum permittivity, and $\epsilon_{SiN}$ is the relative permittivity of SiN. For fundamental transverse-electric-field-like (TE) modes (i.e., the electric field in the plane of the microring), $E_{m,r}$ dominates the integral in the numerator of Eq. (2). Frequency conversion is achieved between the resulting supermodes using only a single monotone and continuous microwave source.

Lithium niobate (LN) is used as the active material to achieve such frequency control in our PhCR, due to its strong nonlinear property, large EO coefficient ($r_{33}$ = 30 pm/V), and wide transparency window (0.4 ~ 5 μm). However, the fabrication of the nano-scale corrugation directly on LN is challenging as LN is a hard material. Any variation of the patterns during LN etching leads to the change of $\beta_m$. Furthermore, the small sidewall angle (40°-80°) of typically reported LNOI (LN-on-insulator) waveguides [19,20] leads to a large deviation between the optical electric field polarization (fundamental TE mode) and the normal direction to the sidewall. This limits $\beta_m$ due to a lower geometry-induced mode coupling strength. To avoid the need for etching LN while using its optical properties, a hybrid SiN-on-LNOI platform (where only SiN but no LN is etched) is a better choice for patterning the PhCR using a CMOS-compatible fabrication process with nano-scale geometry patterning, waveguide

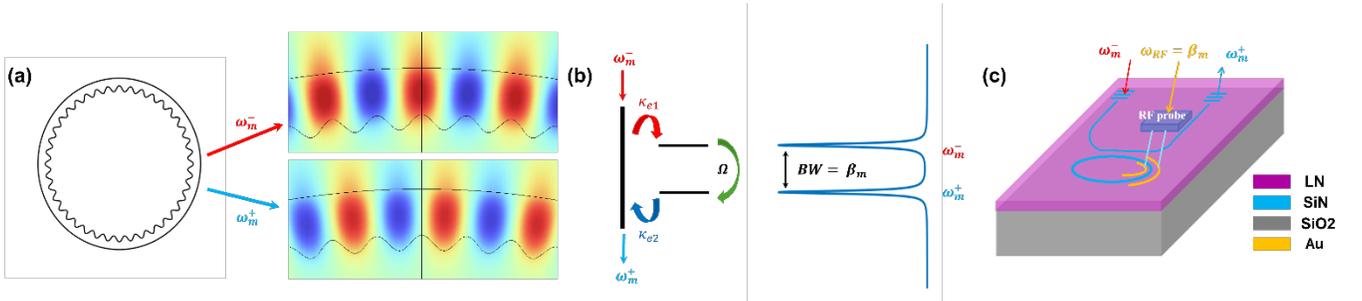

Fig. 1. Schematic of the PhCR and working principle of the PhCR-based frequency converter. (a) A PhCR with periodic geometric corrugation on the inner surface of the circumference of a microring resonator, leading to mode splitting with splitting bandwidth $\beta_m$. (b) The PhCR in (a) can convert the frequency of light from one of the split resonance modes to the other with the coherent microwave excitation at the frequency difference of the two split modes. The conversion is bidirectional with conversion rate related to the coupling rates of the two split resonance modes to the optical waveguide (represented by $\kappa_{e1}$ and $\kappa_{e2}$) and microwave coupling rate $\Omega$. (c) Schematic of the frequency converter in the SiN-on-LNOI platform with grating coupling for the optical input and output. The microwave signal is added with an RF probe.

with nearly vertical sidewalls, and high thermal stability.

In this letter, we demonstrate, for the first time, a precise control of the mode splitting for bidirectional frequency conversion in a PhCR based on a hybrid SiN-on-LNOI platform. The material platform is formed by depositing SiN on an X-cut LNOI substrate using plasma-enhanced chemical vapor deposition (PECVD). All fabrication processes are performed on the SiN layer (no LN etching). A carefully designed PhCR with a reasonably high Q = 1.47 × 10$^5$ and a large splitting bandwidth = 14.6 GHz is demonstrated along with the possibility of tuning the split modes with a voltage-driven shift of 0.85 pm/V. We also demonstrate the precise control of the splitting bandwidth (93.4 MHz/nm) through engineering the geometric structure of the PhCR. These results prove the unique capabilities of our PhCR platform for highly controllable, low-loss, and bidirectional EO frequency conversion.

Figure 1 depicts the operation principle of the PhCR-based microwave-assisted frequency converter. The periodic corrugation on the periphery of the microring resonator splits the two degenerate (CW and CCW) modes of the unperturbed resonator, as schematically shown in Figs. 1(a) and 1(b). Figure 1(c) shows the architecture used for the implementation of the frequency converter in which an optical waveguide is used to couple the input light (e.g., with laser pump frequency $\omega_L = \omega_m^-$) into the PhCR and carry the upconverted signal (e.g., with frequency $\omega_m^+$) from the resonator under the presence of a microwave signal with frequency $\beta_m = \omega_m^+ - \omega_m^-$ (see Fig. 1(b)). The coupling rate between the waveguide mode and the PhCR split modes of frequencies $\omega_m^-$ and $\omega_m^+$ are $\kappa_{e1}$ and $\kappa_{e2}$, respectively. The intrinsic loss rates of the split modes are $\kappa_{i1}$ and $\kappa_{i2}$, respectively. Under the microwave-photon interaction with the optical modes of the PhCR, photons convert from frequency (e.g., $\omega_m^-$) to another (e.g., $\omega_m^+$) with an energy coupling rate of $\Omega$ (Fig. 1(b)). The wideband nature of the waveguide mode allows for a reverse conversion process where input photons of frequency $\omega_m^+$ convert to those of frequency $\omega_m^-$ with the microwave coupling at frequency $\beta_m$. The direction and strength of the frequency conversion are determined by the pump frequency of the input signal in the waveguide ($\omega_L$), $\kappa_{e1}$, $\kappa_{e2}$, and $\Omega$. The conversion efficiency $\eta$ is defined as the ratio of the output power at the converted frequency and the total input power in the waveguide ($\eta = \frac{P_{convert}}{P_{in}}$). The largest frequency conversion efficiency occurs in the over-coupling condition between the waveguide and the resonator with: 1) zero optical detuning between laser pump frequency and the split mode's resonance ($\omega_L = \omega_m^-$ for upconversion), 2) zero microwave detuning between the driven microwave frequency and the optical splitting bandwidth ($\omega_{RF} = \beta_m$), and 3) a matched power coupling rate ($\Omega = \kappa_{e1} + \kappa_{i1}$ for upconversion) [16]. Note that the input/output coupling of the optical signal into/out of the on-chip waveguide is performed using grating couplers (GCs, fabricated in the SiN layer of the hybrid platform), and the microwave excitation is achieved by applying the signal to the LN layer using gold (Au) electrodes as shown in Fig. 1(c).

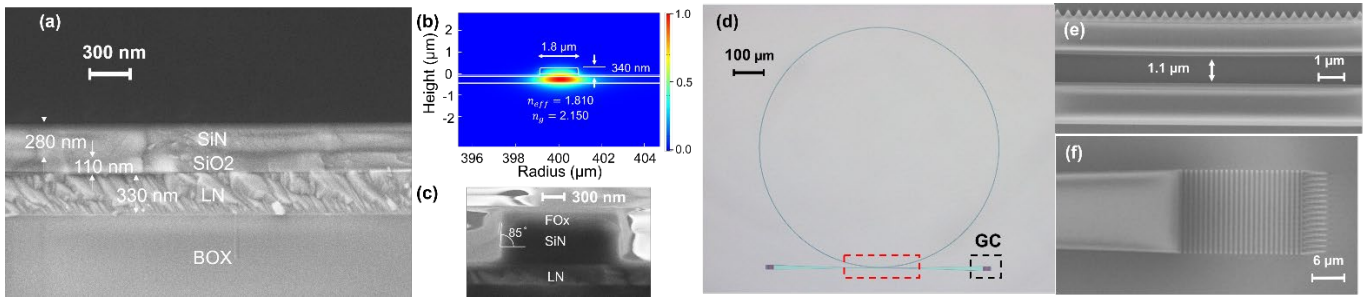

Fig. 2. (a) The SEM image of the hybrid platform, and (b) the axis-symmetric simulation of the fundamental TE mode using COMSOL Multiphysics. (c) The SEM image of the waveguide cross-section showing a sidewall angle of 85°. (d) A top-view microscopy image of the PhCR with waveguide width = 1.8 μm, radius = 400 μm, corrugation amplitude A = 150 nm, and the number of corrugation period n = 5678; the highlighted regions refer to (e) the coupling region between a straight waveguide and a geometric-modulated bend, and (f) the grating coupler for input and output.

Figure 2(a) shows the cross-section scanning electron microscopy (SEM) image of the hybrid platform. The thin-film LN (from NANOLN[TM]) consists of a ~ 330 nm-thick X-cut LN layer on a 4.7 µm buried oxide (BOX, made of silicon oxide, $SiO_2$) layer over a silicon (Si) substrate (525 µm thick). The hybrid platform is formed by depositing a thin (110 nm) film of $SiO_2$ as an etch-stop layer on the LNOI substrate, followed by deposition of a 280 nm-thick SiN layer, both using PECVD. The passive photonic devices are fabricated by first spin-coating flowable oxide (Fox-16) as the electron-beam resist, patterned by electron-beam lithography (EBL). The patterns are fabricated only on the SiN layer with inductively coupled plasma (ICP) etching. The width and radius of the PhCR are 1.8 µm and 400 µm, respectively. Figure 2(b) shows the corresponding simulated electric field profiles of the TE mode. The fabricated waveguides have smooth and nearly vertical sidewalls with a sidewall angle of 85° (Fig. 2(c)), which are critical to get a large splitting bandwidth in PhCRs. The partial existence of the mode in LN allows for EO tuning. Figure 2(d) shows the microscopy image of the PhCR with SEM images of a coupled waveguide (Fig. 2(e)) and input/output GCs (Fig. 2(f)). We fabricate metal electrodes on the LN layer (for DC electric field-driven EO tuning) by spin-coating PMMA (A6) as the second-layer electron-beam resist, patterning by EBL, resist development, metal evaporation (300 nm Au / 15 nm titanium (Ti)), and lift-off in acetone.

The fabricated PhCR shown in Fig. 2(d) is designed with geometric corrugation amplitude $A$ = 150 nm, and number of corrugation period $n$ = 5678. The device is characterized by input/output light coupling between a cleaved optical fiber and the on-chip waveguide using input/output GCs with a tunable laser as input and a detector at the output. Figure 3(a) shows the spectrum of the fundamental TE resonant modes of the PhCR around 1620 nm wavelength with a wide splitting bandwidth = 14.6 GHz while maintaining a high intrinsic quality factor ($Q_{int}$) = $1.47 \times 10^5$. This is very close to $Q_{int}$ = $1.69 \times 10^5$ of the TE resonant mode in a similar microring resonator with no corrugation, fabricated on the same chip (Fig. 3(b)). The resonance is characterized by measuring the transmission through the waveguide in Fig. 2 as a function of wavelength. Figure 3 proves that periodic corrugation in the PhCR introduces strong mode splitting with minimal added loss, which is essential for frequency conversion and other nonlinear applications.

Tunability of the frequency converter is achieved by adding an EO phase shifter to the PhCR resonator as shown in Fig. 4(a). The electric field formed by applying the microwave signal to the electrodes in Fig. 4(a) interacts with the optical mode in the LN layer for EO phase tuning. To maximize the EO effect, a large mode confinement in the LN layer (while maintaining a reasonable $Q_{int}$) is desired. The confinement factor $\Gamma_{LN}$, which represents the ratio of the fundamental TE mode power in the LN layer, can be controlled by the waveguide width and SiN thickness. The detailed design is performed by mode analysis using COMSOL Multiphysics. By using a waveguide width of 1.4 µm for the microring-based phase shifter, we achieve $\Gamma_{LN}$ = 62.5%. The parallel electrodes are patterned with 7.2 µm spacing along the c-axis of the LN crystal to utilize the largest EO coefficient ($r_{33}$ = 30 pm/V) of LN (see Fig. 4(a)). To demonstrate EO tuning, we apply a DC voltage ranging from -20 V to 30 V to the electrodes. Figure 4(b) shows the shift in the resonance wavelength of the PhCR for different applied voltages. It is observed from Fig. 4(b) that $\beta_m$ is relatively unaffected by the applied voltage. Figure 4(c) shows the variation of the resonance shift with the applied voltage, demonstrating a shift ratio of ~ 0.85 pm/V for the split modes (calculated by the average shift of the two split resonances). Figure 4 clearly shows the possibility of tuning the operation frequency of the frequency converter for any desired input signal using an applied voltage.

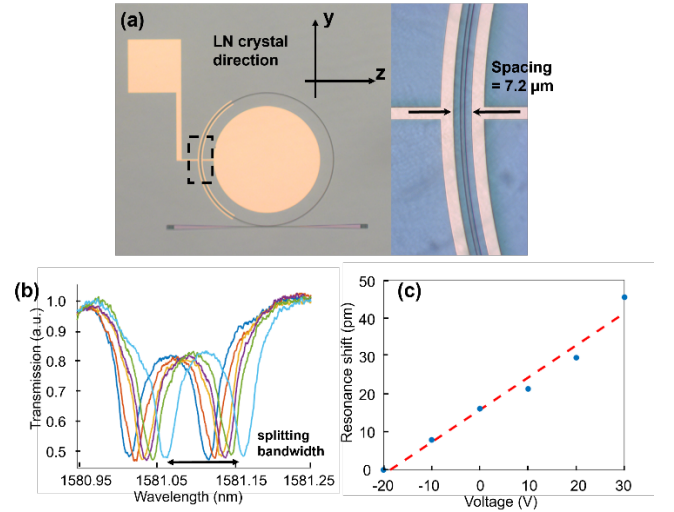

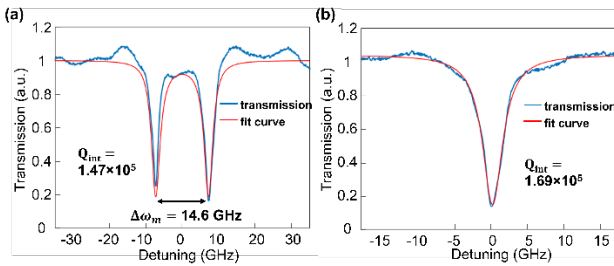

Fig. 3. The normalized optical transmission spectrum and the fitting curve for the measured resonant modes of the fabricated PhCR in Fig. 2(d) with waveguide width = 1.8 µm, bending radius = 400 µm, A = 150 nm, and n = 5678. (a) The split mode in PhCR with $\beta_m$ = 14.6 GHz and $Q_{int}$ = $1.47 \times 10^5$. (b) The single resonant mode in a conventional microring resonators with similar dimensions and A = 0 (no corrugation) showing $Q_{int}$ = $1.69 \times 10^5$. This clearly shows the small effect of the corrugation on $Q_{int}$.

Fig. 4. (a) Microscopy image of the PhCR with radius = 300 µm and width of SiN waveguide = 1.4 µm. The zoom-in image shows two parallel bent electrodes with metal spacing = 7.2 µm. (b-c) Characterization results show a shift of split modes with tunability of 0.85 pm/V and no significant change of $\beta_m$.

One advantage of our PhCRs is precise controllability of the splitting bandwidth using A. Figures 5(a-c) show the SEM images of a portion of PhCRs with same ring dimensions (e.g., waveguide width = 1.4 µm and bending radius = 250 µm) but different values of A. Figures 5(d-g) show the spectra around the resonance wavelength of the same mode with azimuthal number m = 2106 for different values of A. Figure 5(h) clearly shows the linear variation of $\beta_m$ versus A with a fitted slope of 93.4 MHz/nm, thus allowing to carefully design the PhCR for a desired frequency conversion. This linear variation is similar to the reported results for Si [17] and SiN platforms [21]. Such precise control of splitting bandwidth and the large amount of mode splitting ~ 14 GHz (A = 150 nm) are sufficient to achieve bidirectional frequency converters for most practical purposes. To reach larger $\beta_m$ (and thus, larger frequency shifts in the conversion process), larger values of A or smaller widths for the waveguide can be used. The latter results in a weaker confinement of the resonator mode and a stronger field near the periphery of the ring, and thus, a stronger interaction with the corrugated sidewall of the PhCR. The broad normalized transmission of the PhCR with A = 150 nm (with regard to that in Fig. 5(g)) can be found in Supplement 1 (Fig. S1) where we observe a range of split modes due to the anisotropic characteristic of the X-cut LN layer. To pursue single mode splitting with splitting control beyond near-periodic spectrum distribution, the variation of mode indices of the PhCR needs to be compensated. Gradient design of the PhCR is introduced with simulation, fabrication, and characterization results in Supplement 1 (Fig. S2-S3), where the transmission spectrum shows only a few split modes in a much narrower spectrum region. Supplement 1 also shows the calculated near-unity conversion efficiency in an over-coupled waveguide and microresonator.

To summarize, we demonstrated here a new platform for achieving tunable optical frequency converters with precise control of the operation frequency and the dynamic tunability of the frequency shift using a PhCR in the SiN-on-LN platform. Through a CMOS-compatible fabrication process without directly etching LN, we achieved symmetric split modes of a PhCR with a reasonably large $Q_{int} = 1.47 \times 10^5$, a wide splitting bandwidth of 14.6 GHz, and a tuning rate of 0.85 pm/V for the operation frequency of the frequency converter with an applied voltage. A unique advantage of the PhCR technology is the fine controllability of its splitting bandwidth by tuning the corrugation parameter A. We demonstrated the linear relationship between the splitting bandwidth and A with a slope of 93.4 MHz/nm. The main advantages of the PhCR architecture in the SiN-on-LN platform for optical frequency conversion are: simple configuration, large frequency conversion bandwidth (large splitting bandwidth), bidirectional conversion, controllable operation frequency, and CMOS fabrication (no need to etch LN). Beyond frequency conversion, the demonstrated PhCR in the hybrid SiN-on-LNOI platform can contribute to high-performance nonlinear photonic applications such as EO beam splitters and cascaded frequency converters [16].

**Funding.** Army Research Office (GR00008626, Dr. M. Gerhold).

**Disclosures.** The authors declare no conflicts of interest.

**Data availability.** Data underlying the results presented in this paper may be obtained from the authors upon reasonable request.

**Supplemental document.** See Supplement 1 for supporting content.

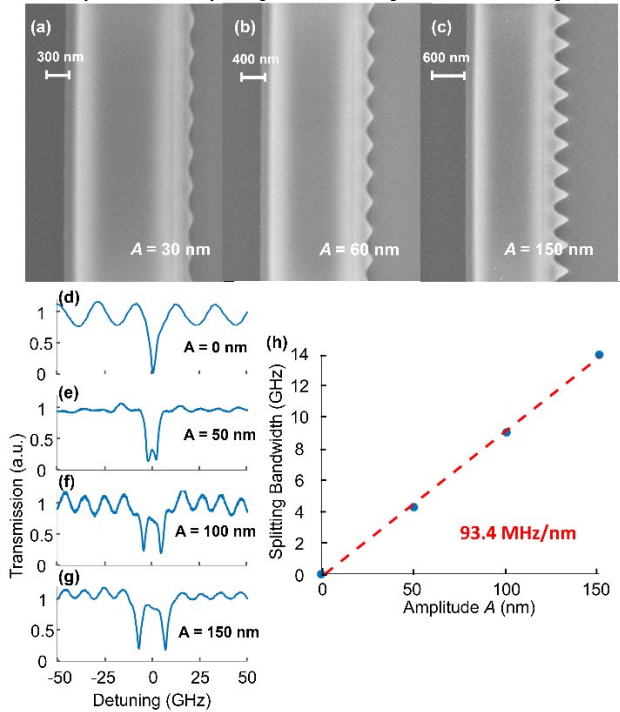

Fig. 5. The precise control of splitting bandwidth in PhCR resonators with waveguide width = 1.4 μm and bending radius = 250 μm. (a-c) The SEM images of the bent waveguides in PhCR with A= 30 nm, 60 nm, and 150 nm, respectively. Normalized transmission spectra of the resonant modes in (d) a simple ring resonator (A = 0) and (e-g) the PhCRs with various $A$ = 50, 100, and 150 nm, respectively. The resonant modes with wavelength around 1620 nm have the same azimuthal mode number ($m$ = 2106) in all cases. (h) Mode splitting as a function of $A$ in different PhCRs.

## REFERENCES

1. H. Feng, T. Ge, X. Guo, et al., Nature **627**, 80 (2024).
2. D. Marpaung, J. Yao, and J. Capmany, Nature photonics **13**, 80 (2019).
3. J. M. Lukens, H.-H. Lu, B. Qi, et al., Journal of Lightwave Technology **38**, 1678 (2019).
4. A. Yi, C. Wang, L. Zhou, et al., Applied Physics Reviews **9**(2022).
5. T. Kobayashi, R. Ikuta, S. Yasui, et al., Nature photonics **10**, 441 (2016).
6. Q. Li, M. Davanço, and K. Srinivasan, Nature Photonics **10**, 406 (2016).
7. L. Shao, N. Sinclair, J. Leatham, et al., Optics Express **28**, 23728 (2020).
8. D. B. Sohn, S. Kim, and G. Bahl, Nature Photonics **12**, 91 (2018).
9. L. J. Wright, M. Karpiński, C. Söller, et al., Physical review letters **118**, 023601 (2017).
10. L. Fan, C.-L. Zou, M. Poot, et al., Nature Photonics **10**, 766 (2016).
11. X. He, L. Cortes-Herrera, K. Opong-Mensah, et al., Optics Letters **47**, 5849 (2022).
12. S. F. Preble, Q. Xu, and M. Lipson, Nature Photonics **1**, 293 (2007).
13. H.-P. Lo and H. Takesue, Optica **4**, 919 (2017).
14. A. Savchenkov, W. Liang, A. Matsko, et al., Optics Letters **34**, 1300 (2009).
15. Y. Zheng, A. Yi, C. Ye, et al., in 2022 Conference on Lasers and Electro-Optics (CLEO), (IEEE, 2022).
16. Y. Hu, M. Yu, D. Zhu, et al., Nature **599**, 587 (2021).
17. X. Lu, Silicon and Silicon Carbide Photonics and the Applications (University of Rochester, 2016).
18. Z. Peng, R. Krishna, X. Wu, et al., in 2024 Conference on Lasers and Electro-Optics (CLEO), (IEEE, 2024).
19. D. Zhu, L. Shao, M. Yu, et al., Advances in Optics and Photonics **13**, 242 (2021).
20. A. Kozlov, D. Moskalev, U. Salgaeva, et al., Applied Sciences **13**, 2097 (2023).
21. X. Lu, A. Rao, G. Moille, et al., Photonics research **8**, 1676 (2020).